\documentclass[%
 aip,
 pof,%
 amsmath,amssymb,
reprint,%
onecolumn
]{revtex4-1}
\usepackage {CJK}
\usepackage{graphicx}
\usepackage{dcolumn}
\usepackage{bm}
\usepackage[mathlines]{lineno}
\usepackage{color}
\usepackage{hyperref}

\hypersetup{
}

\usepackage{makecell}
\begin{document}

\title{Wall confinement effects on the dynamics of cavitation bubbles in thin tubes}
\author{Nian Wang}
\affiliation{State Key Laboratory of Engines, Tianjin University, Tianjin, 300350, China.}%
\author{Huashi Xu}
\affiliation{State Key Laboratory of Engines, Tianjin University, Tianjin, 300350, China.}%
\author{Tianyou Wang}
\affiliation{State Key Laboratory of Engines, Tianjin University, Tianjin, 300350, China.}%
\affiliation{National Industry-Education Platform of Energy Storage, Tianjin University, Tianjin, 300350, China}
\author{Zhizhao Che}
\email{chezhizhao@tju.edu.cn}
 \affiliation{State Key Laboratory of Engines, Tianjin University, Tianjin, 300350, China.}
 \affiliation{National Industry-Education Platform of Energy Storage, Tianjin University, Tianjin, 300350, China}
\date{\today}

\begin{abstract}
Cavitation is a common phenomenon in nature and has numerous applications. In contrast to a cavitation bubble in a free domain, a cavitation bubble in a thin tube is restricted by the tube wall, which is expected to significantly affect bubble evolution but its mechanism is still unclear. In this study, the dynamics of a cavitation bubble in a thin circular tube is studied by numerical simulation, focusing on the confinement effects of the tube. The results show that besides affecting the size and lifetime of the bubble, the confinement effects of the tube lead to the generation of counter jets and a ring jet during the contraction process of the bubble, and the curvature of the two counter jets determines the ring jet's peak velocity. When the bubble deviates from the midpoint of the tube in the axial direction, the two sides of the bubble along the axial direction show asymmetric behaviors, which results in the bubble migrating toward the midpoint. The tube diameter, tube length, liquid viscosity, and initial bubble position, can significantly influence the degree of confinement effects, which can be characterized by the variations of several key indicators, such as bubble size, lifetime, degree of deformation, counter jet velocity, ring jet velocity, and axial migration of the bubble.
\end{abstract}


\maketitle

\section{Introduction}
Cavitation is a ubiquitous phenomenon in nature and engineering applications, such as propellers \cite{subhas12}, pumps \cite{jia23}, nozzles \cite{sou07}, material synthesis \cite{xu13}, surface cleaning \cite{chahine16}, and biomedicine \cite{brennen15}. Cavitation occurs as a localized vaporization phenomenon due to the liquid pressure decreasing below the saturated vapor pressure. Since the collapse time of a cavitation bubble is very short, a huge energy pulse is generated at the instant of collapse, which leads to a series of effects on the surroundings, such as cavitation erosion \cite{bilus20}, noise \cite{si23}, and sonoluminescence \cite{suslick08}.

In a free domain of liquids, a cavitation bubble expands and contracts in a spherical shape. The evolution of a spherical bubble in a free domain can be described by the Rayleigh equation \cite{rayleigh17} if neglecting the fluid viscosity and surface tension. Plesset \cite{plesset49} derived the Rayleigh-Plesset equation based on the Rayleigh equation by adding the influences of surface tension, liquid viscosity, and non-condensable gases inside the bubble on bubble motion. In the final stage of collapse, the bubble evolution is very rapid, generating very high pressure and temperature at the location of collapse \cite{qin16} and resulting in a bright light \cite{schanz12}. The bubble generates shock waves during collapse and then rebounds \cite{akhatov01}. The bubble fissions when it collapses to the minimum volume, thus deviating from the spherical, and the energy dissipation caused by the fission results in a decrease in the volume of the rebounding bubble \cite{brennen02}. In addition, the interaction of two bubbles in the free domain generates liquid jets directed toward or away from each other \cite{han15}.

Cavitation also often occurs in confined spaces in many applications, such as fluid engineering \cite{ji15, li00, wang18}, microfluidics \cite{marmottant04, zwaan07, shafaghi21}, and clinical medicine \cite{chen11, hosny13}. Compared with free domains, cavitation bubble dynamics in confined spaces is more complex, including the evolvement of the bubble shape and the development of liquid jets \cite{cui20, cui23, gonzalez-avila20, zhang21}. Different types of confinements were considered for cavitation bubbles, such as rigid walls \cite{wang20, xu23}, free surface \cite{li19interactions, zhang18}, elastic boundary \cite{brujan01, reese23}, and mixed boundaries \cite{cui16, li19, zhang15, li23}, which interact with the bubble and result in complex bubble dynamics, such as changes in the shape and lifetime of the bubble, generation of liquid jets, and migration of the bubble.

Cavitation in a tube is a common cavitation situation in confined domains. With the confinement effects of the tube wall, the lifetime of the bubble is significantly increased \cite{aghdam15}. During the bubble contraction process, two counter jets point to the tube midpoint in the axial direction, and then generate a ring jet to impact the tube wall \cite{ni12}. When the bubble deviates from the midpoint of the tube in the axial direction, the pumping effect generated by the bubble evolution causes the liquid in the tube to flow from one side to the other \cite{yuan99, ory00}, and secondary cavitation is generated at the time of the bubble collapsing \cite{ji17}. When the bubble deviates from the midpoint of the tube in the radial direction, two different jets are generated, directing towards the closer or farther wall, respectively \cite{wang19}. When two bubbles are in the tube, the interaction of bubble pairs leads to more significant deformation of the bubbles, and the liquid flow in the tube becomes more complex after the bubbles collapse \cite{zou15}. The dynamics of bubbles in the profiled tube are more complex, including asymmetric liquid jets \cite{ren22} and splitting of the bubbles \cite{nagargoje23}.

Although cavitation bubbles in tubes have been studied by previous researchers, the understanding of the process is still insufficient. Particularly, the confinement effects of the wall on the bubble dynamics are still unclear, such as the generation of liquid jets and the migration of the bubble. Hence, the evolution of a cavitation bubble in a thin circular tube is investigated by numerical simulation in this study. The velocity and pressure distribution during bubble evolution are analyzed, and the influences of key parameters, such as tube diameter, tube length, and initial bubble position, on the cavitation process are investigated, focusing on the important characteristics of confined cavitation bubbles such as the generation of counter jets and ring jets, and the migration of the bubble.

\section{Numerical methods}

\subsection{Governing equations}\label{sec:2.1}
The simulation was performed in OpenFOAM using the pressure-based compressible two-phase flow solver compressibleInterFoam \cite{weller98}. The finite volume method (FVM) was used for solving the compressible Navier-Stokes equations, and the volume-of-fluid (VOF) method for capturing the gas-liquid interface. The conservation equations for mass, momentum, energy, and volume fraction are respectively:
\begin{equation}\label{eq:01}
\frac{\partial \rho }{\partial t}+\nabla \cdot (\rho \mathbf{u})=0,
\end{equation}
\begin{equation}\label{eq:02}
\frac{\partial (\rho \mathbf{u})}{\partial t}+\nabla \cdot (\rho \mathbf{uu})=-\nabla p+\nabla \cdot \tau +{\mathbf{f}_{\sigma }},
\end{equation}
\begin{equation}\label{eq:03}
\frac{\partial (\rho T)}{\partial t}+\nabla \cdot (\rho \mathbf{u} T)+\frac{1}{{{c}_{v}}}\left( \frac{\partial (\rho k)}{\partial t}+\nabla \cdot (\rho \mathbf{u} k)+\nabla \cdot (p \mathbf{u}) \right) =\nabla \cdot ({{\alpha }_\text{eff}}\nabla T),
\end{equation}
\begin{equation}\label{eq:04}
\frac{\partial {{\alpha }_{l}}}{\partial t}+\nabla \cdot ({{\alpha }_{l}}\mathbf{u})+\nabla \cdot ({{\alpha }_{l}}{{\alpha }_{v}}{\mathbf{u}_{r}})={{\alpha }_{l}}{{\alpha }_{v}}(\frac{1}{{{\rho }_{v}}}\frac{d{{\rho }_{v}}}{dt}-\frac{1}{{{\rho }_{l}}}\frac{d{{\rho }_{l}}}{dt})  +{{\alpha }_{l}}\nabla \cdot \mathbf{u},
\end{equation}
where $\mathbf{u}$ is the velocity vector, $\rho$ is the fluid density, $p$ is the pressure, $\tau = \mu \left( \nabla \mathbf{u} +\nabla {{\mathbf{u}}^\mathsf{T}}-2/3(\nabla \cdot \mathbf{u})\mathbf{I} \right)$ is the viscous stress tensor, $\mu$ is the dynamic viscosity, $\mathbf{I}$ is the unit tensor, $\mathbf{f}_\sigma$ is the source term due to surface tension calculated by the continuum surface force (CSF) model \cite{brackbill92}. $T$ is the temperature, $c_v$ is the specific heat capacity, $\alpha_\text{eff}$ is thermal diffusivity, $k=0.5{{\left| \mathbf{u} \right|}^{2}}$ is the specific kinetic energy, $\alpha_i$ is the phase volume fraction, and $\mathbf{u}_r$ is the relative velocity between the two phases. The subscripts $l$ and $v$ indicate the liquid and gas phases.

The physical properties were calculated from the volume fraction:
\begin{equation}\label{eq:05}
\beta ={{\alpha }_{l}}{{\beta }_{l}}+{{\alpha }_{v}}{{\beta }_{v}},
\end{equation}
where $\beta_l$ and $\beta_v$ denote the properties of the liquid and gas phases, such as the density $\rho$ and dynamic viscosity $\mu$.

The Tait equation of state \cite{koch16} was used for the liquid
\begin{equation}\label{eq:06}
{\rho}_{l} ={{\rho }_{\text{ref}}}{{\left( \frac{p+B}{{{p}_{\text{ref}}}+B} \right)}^{\frac{1}{{{n}_{l}}}}},
\end{equation}
where $\rho_\text{ref}$ and $p_\text{ref}$ are the reference density and pressure, which are 998 kg/m$^3$ and 101325 Pa, respectively, and $n_l$ and $B$ are 7.15 and $3.046\times {{10}^{8}}$ Pa, respectively.

The ideal gas equation of state was used for the gas
\begin{equation}\label{eq:07}
{{\rho }_{v}}=\frac{p}{RT}.
\end{equation}

\subsection{Numerical implementation}\label{sec:2.2}

\begin{figure}
  \centering
  \includegraphics[scale=0.9]{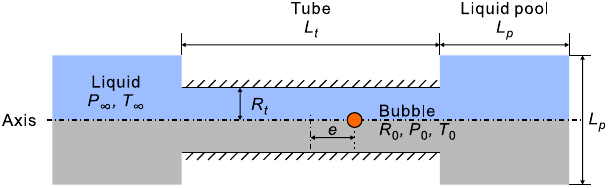}
  \caption{Schematic diagram of the numerical setting for a cavitation bubble in a tube.}
  \label{fig:01}
\end{figure}

The simulation setup follows the cavitation bubble experiments in the literature \cite{wang19}. To reduce the computational cost, an axisymmetric setup was used, as shown in Figure \ref{fig:01}. The length of the tube is $L_t$, the radius is $R_t$, and the tube wall was set with an adiabatic no-slip boundary condition. Large cylindrical liquid pools ($L_p = 8 R_t$ in both diameter and height) were set at the two sides of the tube, and the boundaries of the pools were set as free inlet/outlet.

In the simulation, the initial bubble was assumed to be spherical with high internal pressure and zero velocity, with initial radius, pressure, and temperature as $R_0$, $P_0$, and $T_0$, respectively. The center of the bubble was placed on the axis of the tube with an axial offset distance $e$ from the midpoint of the tube. The initial pressure and temperature of the liquid in the tube and the pool were set to $P_\infty$ and $T_\infty$.

The theoretical maximum radius $R_m$ and maximum velocity $v_m$ of the bubble were obtained by solving the Rayleigh equation \cite{rayleigh17}
\begin{equation}\label{eq:08}
1-\frac{R_{m}^{3}}{R_{0}^{3}}+\frac{{{P}_{0}}}{{{P}_{\infty }}}\ln \left( \frac{R_{m}^{3}}{R_{0}^{3}} \right)=0,
\end{equation}
\begin{equation}\label{eq:09}
{{v}_{m}}^{2}=\frac{2{{P}_{0}}}{3{{\rho }_{l}}}\exp \left( \frac{{{P}_{\infty }}-{{P}_{0}}}{{{P}_{0}}} \right)-\frac{2{{P}_{\infty }}}{3{{\rho }_{l}}}.
\end{equation}

To analyze the confinement effect of the tube, several dimensionless parameters affecting the bubble dynamics were defined. The dimensionless tube diameter and length are
\begin{equation}\label{eq:10}
{{R}_{t}}^{*}=\frac{{{R}_{t}}}{{{R}_{m}}},
\end{equation}
\begin{equation}\label{eq:11}
{{L}_{t}}^{*}=\frac{{{L}_{t}}}{{{R}_{m}}}.
\end{equation}
The dimensionless initial offset of the bubble in the tube from the mid-point is
\begin{equation}\label{eq:12}
\gamma =\frac{e}{{{R}_{m}}}.
\end{equation}
The dimensionless time is
\begin{equation}\label{eq:13}
{{t}^{*}}=\frac{t}{{{t}_{m}}},
\end{equation}
where
\begin{equation}\label{eq:14}
{{t}_{m}}={{R}_{m}}\sqrt{\frac{{{\rho }_{l}}}{{{P}_{\infty }}}}.
\end{equation}

\subsection{Grid independence study}
To check the accuracy of the results, a grid independence study and an experimental validation were carried out. In the simulation, the bubble was set to be spherical at the midpoint of the tube initially, with ${{R}_{0}}=2$ mm, ${{P}_{0}}=1.3\times {{10}^{7}}$ Pa, and ${{T}_{0}}=400$ K. The liquid surrounding pressure and temperature are ${{P}_{\infty }}=101325$ Pa, and ${{T}_{\infty }}=297$ K, respectively. The tube size is consistent with the experimental setup in Ref.\ \citenum{wang19}, i.e., ${{R}_{t}}=10$ mm, ${{L}_{t}}=160$ mm, ${{R}_{m}}=19.2$ mm, ${{v}_{m}}=56$ m/s, and ${{t}_{m}}=1.9$ ms. Their corresponding dimensionless parameters are $R_t^* = 0.521$, $L_t^* = 8.333$, and $\gamma =0$.

\begin{figure}
  \centering
  \includegraphics[scale=0.9]{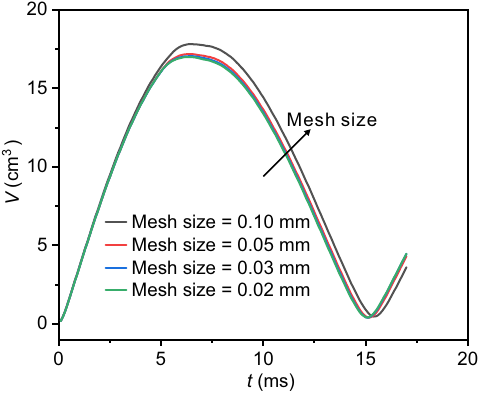}
  \caption{Mesh convergence test for the evolution of the bubble size with different mesh sizes. Here, $R_t^* = 0.521$, $L_t^* = 8.333$, and $\gamma = 0$.}
  \label{fig:02}
\end{figure}

We used four meshing resolutions for the grid independence study. A uniform mesh is used for the tube region and coarser mesh is used in the liquid pool region with a gradual transition. The average mesh size in the pool region is four times that of the tube region. The mesh was varied progressively with mesh sizes of 0.1, 0.05, 0.03, and 0.02 mm, respectively. The results of bubble volume $V$ are shown in Figure \ref{fig:02}. As the mesh size reduces below 0.03 mm, further reducing the mesh size has no significant effect on the bubble volume. Therefore, a mesh size of 0.03 mm was used for the simulation.

\subsection{Experimental validation}
\begin{figure*}
  \centering
  \includegraphics[scale=0.75]{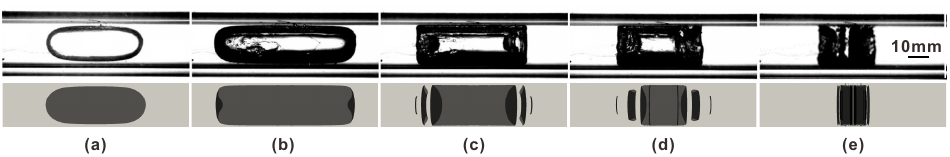}
  \caption{Comparison of numerical simulation results (bottom row) and experimental results in Ref.\ \citenum{wang19} (top row). Here $R_t^* = 0.521$, $L_t^* = 8.333$, $\gamma = 0$. The dimensionless times in the figures are: (a) $\hat{t}=-0.3$, (b) $\hat{t}=0$, (c) $\hat{t}=0.7$, (d) $\hat{t}=0.8$, (e) $\hat{t}=1$. The dimensionless time is defined as $\hat{t}=\left( t-{{t}_{\max }} \right)/\left( {{t}_\text{coll}}-{{t}_{\max }} \right)$, where $t_\text{max}$ is the time of maximum bubble size, $t_\text{coll}$ is the bubble collapse time. In the simulation results, the evolution of the bubble profile is represented by the contour of $\alpha = 0.5$. The dimensional times in the experiment are: 3.12, 5.53, 10.41, 11.71, and 12.94 ms; the dimensional times in the simulation are: 2.5, 5.5, 11.7, 13.2, and 14.9 ms. The slight misalignment in the dimensional times could be because the initial internal pressure and radius of the bubble are unknown and the simulation was initiated by assuming reasonable values. In the experimental results, the center of the bubble is bright and the edge of the bubble is dark; in the simulation results, the bubble interface is dark gray and the depression on the bubble interface is black. The experimental images are reprinted from International Journal of Multiphase Flow, 121, Shi-Ping Wang, Qianxi Wang, A-Man Zhang, Eleanor Stride, Experimental observations of the behaviour of a bubble inside a circular rigid tube, 103096, Copyright (2019), with permission from Elsevier.}
  \label{fig:03}
\end{figure*}

Comparison of the simulation results with the experimental results in Ref.\ \citenum{wang19} is shown in Figure \ref{fig:03}. Compared with the experimental images, the numerical simulation can capture the bubble evolution process well, and the main dynamics during the bubble evolution process are well reproduced.

\section{Results and discussion}
\subsection{Bubble evolution process}\label{sec:3.1}

\begin{figure*}
  \centering
  \includegraphics[scale=0.75]{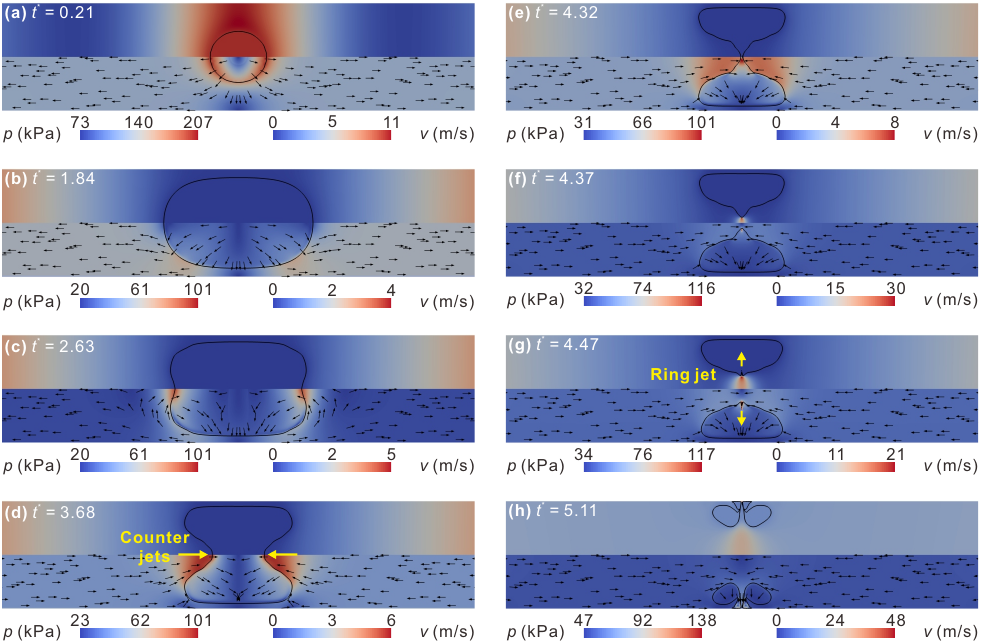}
  \caption{Bubble shape as well as pressure (upper halves of the figures) and velocity distributions (lower halves of the figures) during bubble evolution. Here $R_t^* = 0.781$, $L_t^* = 8.333$, and $\gamma = 0$ (Multimedia available online).}
  \label{fig:04}
\end{figure*}

The bubble evolution under a typical condition ($R_t^* = 0.781$, $L_t^* = 8.333$, $\gamma = 0$) is shown in Figure \ref{fig:04} (Multimedia available online). The bubble expands quickly in the initial stage because of the large internal pressure (see Figure \ref{fig:04}a). Subsequently, the radial expansion of the bubble is restricted by the tube wall, and the gas-liquid interface expands significantly faster along the axial direction of the tube than in the radial direction, resulting in an ellipsoidal shape of the elongated bubble along the axial direction. When the bubble expands close to the tube wall, its radial expansion is restricted by the tube wall, causing the bubble to expand primarily in the axial direction. The pressure inside the bubble gradually decreases to less than that of the surrounding liquid as the bubble expands (see Figure \ref{fig:04}b). After expanding to its maximum size, the bubble begins to contract along the tube axis and forms two counter jets pointing to the midpoint of the tube (see Figures \ref{fig:04}c-d). Because of the resistance of the tube wall to the nearby liquid, the liquid velocity away from the wall is higher than that near the wall, and the liquid velocity at $r = 0$ is the largest. The velocity of the counter jets increases gradually before meeting at the midpoint of the tube (see Figure \ref{fig:04}e). The counter jets meet and penetrate the bubble, producing a donut shape (see Figure \ref{fig:04}f). The counter jets meet and further form a ring jet outward in the radial direction (see Figures \ref{fig:04}g-h).

\begin{figure*}
  \centering
  \includegraphics[scale=0.75]{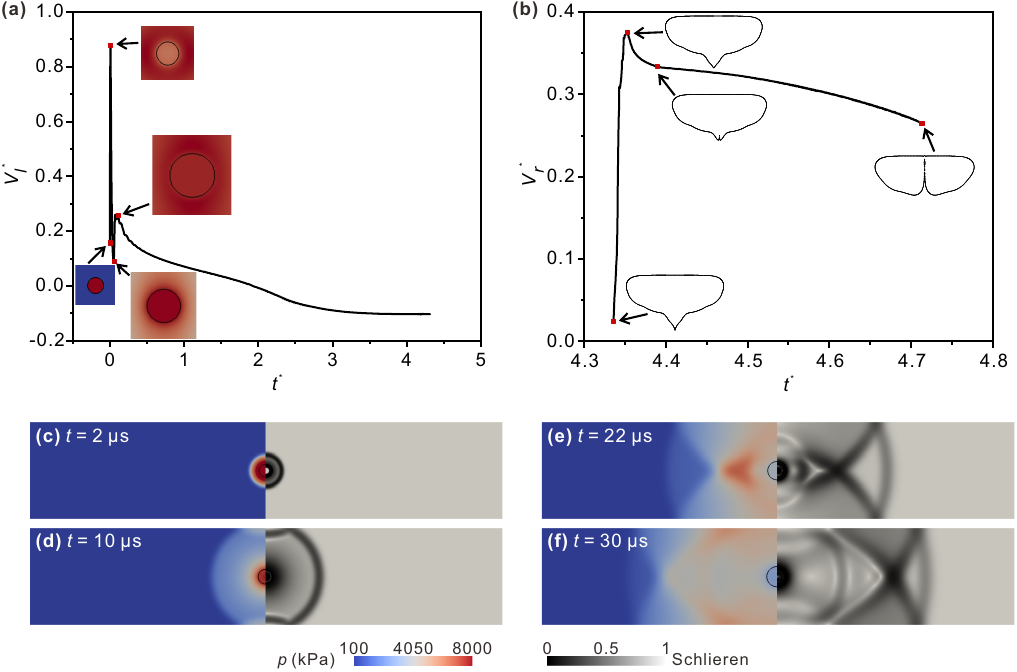}
  \caption{(a-b) Variation of (a) counter jet velocity $v_l^*$ and (b) ring jet velocity $v_r^*$ during bubble evolution. The insets in panel (a) show the pressure distributions, while the insets in panel (b) show the bubble shapes. Here, $R_t^* = 0.781$, $L_t^* = 8.333$, and $\gamma = 0$. (c-f) Pressure wave propagation in the early expansion stage (left halves) and its numerical Schlieren image (right  halves).}
  \label{fig:05}
\end{figure*}

The counter jets and ring jets are important interface phenomena during bubble evolution. To characterize the formation and development of counter jets, we extract the variation of the jet velocity ($v_l$, positive for outwards and negative for inwards) in the axial direction at the gas-liquid interface before the counter jets meet. We define a dimensionless velocity of the counter jets
\begin{equation}\label{eq:15}
v_{l}^{*}=\frac{{{v}_{l}}}{{{v}_{m}}},
\end{equation}
where $v_m$ is the theoretical maximum velocity obtained from Eq.\ (\ref{eq:09}). As shown in Figure \ref{fig:05}a, $v_l^*$ changes with a rapid increase followed by rapid fluctuation and then gradually decreases to negative. This is because in the early stage of bubble expansion, the bubble pressure is remarkably higher than that of the surrounding liquid, and the bubble expands at a high speed driven by the pressure difference, resulting in a large axial velocity $v_l^*$. Subsequently, the bubble interface velocity fluctuates rapidly because of the propagation and reflection of shock waves in the tube. The pressure change caused by the rapid bubble expansion results in the surrounding pressure exceeding the bubble pressure, forming a resistance to the bubble expansion, and leading to a rapid decrease in the velocity of gas-liquid interface motion. When the axial velocity decreases to zero, the bubble begins to contract and $v_l^*$ becomes negative until the counter jets meet.

To characterize the rapid pressure change in the tube in the early period of bubble expansion, the pressure distribution and its numerical Schlieren image are shown in Figures \ref{fig:05}c-f. The Schlieren value was calculated as $\exp \left( -k\left| \nabla \rho  \right|/\max \left| \nabla \rho  \right| \right)$, where $k$ is a free parameter. As shown in Figures \ref{fig:05}c-d, the high-speed expansion of the bubble generates a spherical shockwave propagating outward. The reflected shockwaves meet besides the bubble, generating a high-pressure region, while the shockwave is reflected by the bubble as a rarefaction wave, which propagates outward in the axial direction (see Figure \ref{fig:05}e). As the pressure wave continues to propagate, a complex pressure distribution is produced inside the tube (see Figure \ref{fig:05}f). The propagation and reflection of the pressure wave also caused rapid fluctuation in the velocity of the bubble interface, as shown in Figure \ref{fig:05}a.

To characterize the generation and development of the ring jet, we extract the ring jet velocity (i.e., the radial velocity $v_r$ at the gas-liquid interface of the ring jet) and define the dimensionless ring jet velocity
\begin{equation}\label{eq:16}
v_{r}^{*}=\frac{{{v}_{r}}}{{{v}_{m}}}.
\end{equation}
As shown in Figure \ref{fig:05}b, $v_r^*$ changes with time in a trend of rapid increase and then gradually decreases. This is due to the two counter jets meeting in the tube at the midpoint, which forms a high-pressure region. The pressure gradient in the radial direction leads to the formation of the ring jet, which quickly reaches the maximum speed. Subsequently, the speed of the ring jet decreases and the tip begins to break up under the effect of surface tension, forming small droplets that move towards the tube wall.

\subsection{Effect of tube diameter}\label{sec:3.2}
\begin{figure}
  \centering
  \includegraphics[scale=0.85]{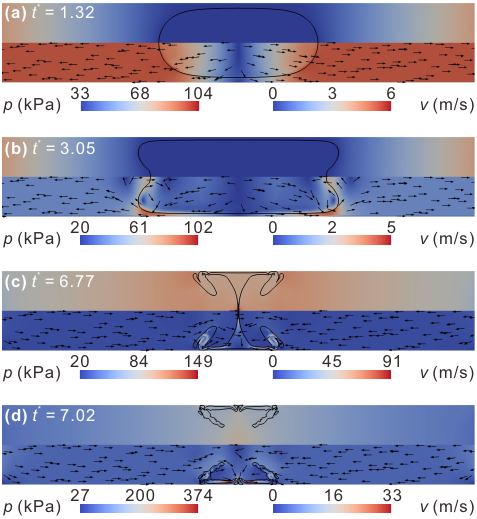}
  \caption{Bubble shape as well as pressure and velocity distributions during bubble evolution for a smaller tube diameter. Here, $R_t^* = 0.573$, $L_t^* = 8.333$, and $\gamma = 0$ (Multimedia available online).}
  \label{fig:06}
\end{figure}

To analyze the influence of the tube diameter on the bubble dynamics, the bubble evolution at a smaller tube diameter ($R_t^* = 0.573$) are shown in Figure \ref{fig:06} (Multimedia available online). Compared with $R_t^* = 0.781$ in Figure \ref{fig:04}, the bubble with a smaller tube diameter has a larger aspect ratio during the expansion process, and the bubble completely occupies the tube cross-section during its evolution, which leads to almost no liquid circulation between the two sides of the bubble (see Figures \ref{fig:06}a-b). Before the counter jets meet, as the bubble interface is closer to the wall than that in larger tubes, the viscous resistance near the wall causes the local contraction speed of the bubble interface to be lower than that on the axis. As a result, a larger curvature is generated at the two sides of the bubble near the wall, and then the bubble splits into small bubbles, as shown in Figure \ref{fig:06}c. In addition, when penetrated by the ring jet, the bubble is further deformed to a greater degree as compared to that in larger tubes (Figure \ref{fig:04}h), as shown in Figure \ref{fig:06}d.

\begin{figure*}
  \centering
  \includegraphics[scale=0.75]{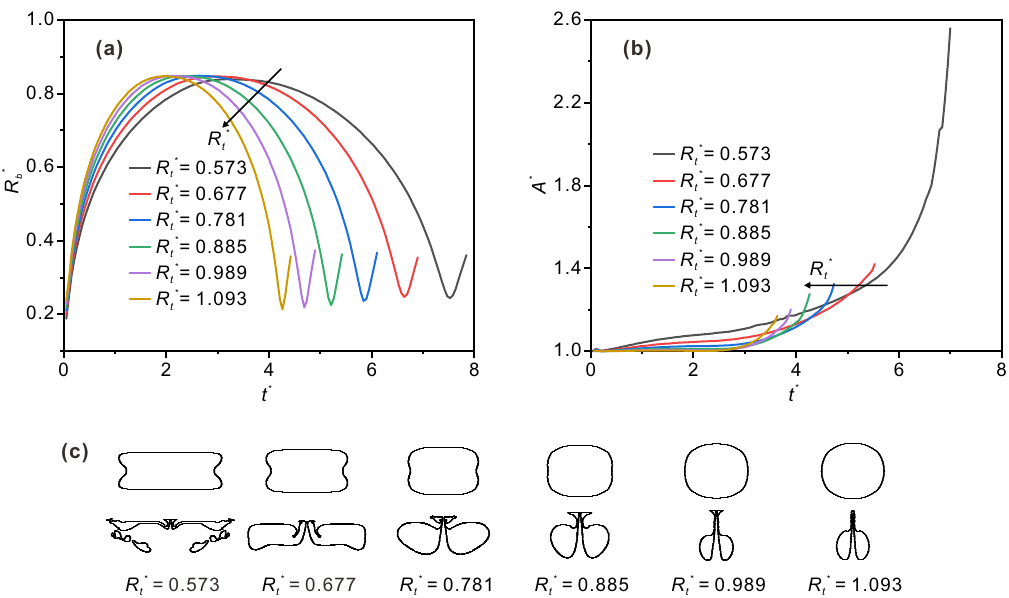}
  \caption{Influence of the tube diameter $R_t^*$ on the bubble shape with  $L_t^*$ = 8.333, $\gamma = 0$: (a) variation of the bubble size $R_b^*$; (b) variation of the deformation parameter $A^*$; (c) bubble shapes at the time of maximum bubble size (the first row) and at the moment of ring jet penetration (the second row).}
  \label{fig:07}
\end{figure*}

To characterize the confinement effect of the tube on bubble expansion, we define a dimensionless size parameter
\begin{equation}\label{eq:17}
{{R}_{b}}^{*}=\frac{\sqrt[3]{3V/4\pi }}{{{R}_{m}}},
\end{equation}
where $V$ is the instantaneous bubble volume. The time variation of $R_b^*$ for different tube diameters is shown in Figure \ref{fig:07}a. With the decrease in $R_t^*$, the confinement effect of the wall increases, the bubble expansion rate slows down, and the bubble lifetime increases. In addition, with the decrease in $R_t^*$, the maximum size of the bubble decreases. This indicates that the confinement effect of the tube wall during bubble evolution causes resistance to bubble expansion, resulting in a smaller maximum size.

To characterize the degree of bubble deformation during bubble evolution, we define a dimensionless deformation parameter
\begin{equation}\label{eq:18}
{{A}^{*}}=\frac{\sqrt{A/4\pi }}{\sqrt[3]{3V/4\pi }},
\end{equation}
where $A$ is the instantaneous surface area of the bubble. Hence, $A^*$ represents the degree to which the bubble deviates from spherical. If the bubble is perfectly spherical, $A^*=1$, and a larger $A^*$ indicates that the bubble deviates more from a spherical shape. The time variation of the dimensionless deformation parameter $A^*$ from $t^* = 0$ to the instant when the ring jet penetrates the bubble for different $R_t^*$ is shown in Figure \ref{fig:07}b. In the whole process, $A^*$ increases with time, which indicates that the degree of bubble deformation gradually increases. $A^*$ increases as $R_t^*$ decreases, this indicates the increased confinement effect of the tube wall leading to the increased degree of bubble deformation.

The bubble shape at the moments of maximum size and the ring jet penetration for different tube diameters $R_t^*$ are shown in Figure \ref{fig:07}c. With the decrease in tube diameter $R_t^*$, the confinement effect of the tube increases, and the bubble shape is further away from spherical. At a small $R_t^*$  (e.g., $R_t^*  = 0.573$), the bubble elongates along the axis and the gas-liquid interface at the axis starts to contract inwards, leading to a local concave shape of the interface. At the instant of ring jet penetration, the bubble shape becomes more irregular with the decrease in tube diameter. This is because of the higher local viscous effect caused by the tube wall in the bubble contraction stage, promoting the bubble deformation, or even resulting in bubble breakup in the case of small $R_t^*$, such as $R_t^* = 0.573$.

\begin{figure*}
  \centering
  \includegraphics[scale=0.75]{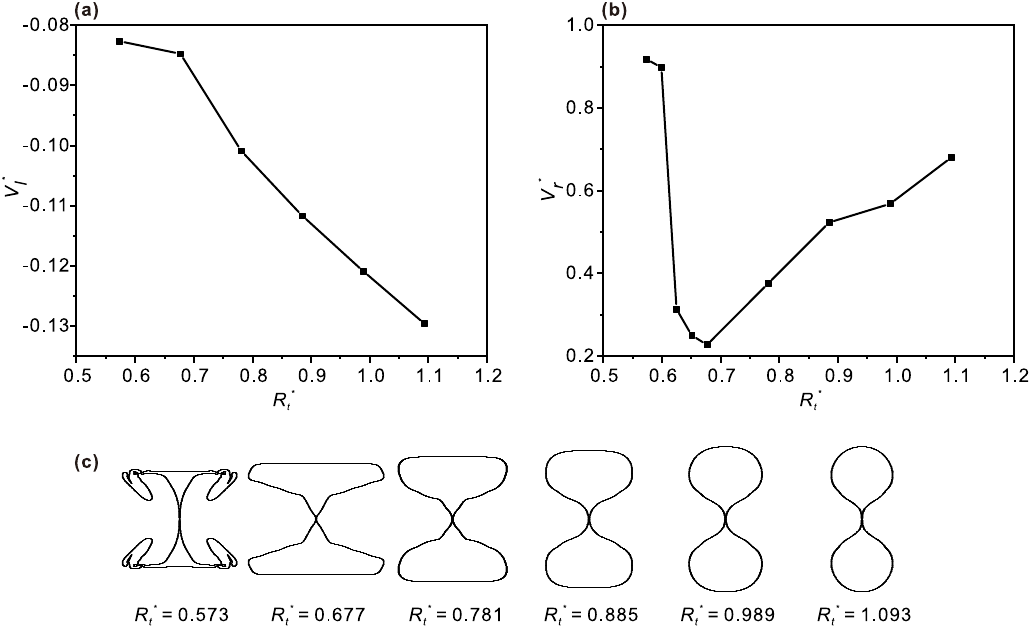}
  \caption{Influence of the tube diameter $R_t^*$ on the counter jets and the ring jet with $L_t^*$ = 8.333, $\gamma = 0$: (a) jet velocity $v_l^*$ at the meeting moment of the counter jets; (b) peak velocity $v_r^*$ of the ring jet; (c) bubble shapes at the meeting moment of the counter jets. }
  \label{fig:08}
\end{figure*}

Now we focus on the influence of the tube diameter on the counter jets and the ring jet. The jet velocity at the meeting time of the two counter jets for different tube diameters $R_t^*$ is shown in Figure \ref{fig:08}a. With the decrease in $R_t^*$, the magnitude of $v_l^*$ decreases (note that $v_l^*$ is negative in Figure \ref{fig:08}a). This is mainly due to the increased restriction by the tube wall, resulting in a suppression of the bubble's contraction velocity. The peak velocity of the ring jet for different $R_t^*$ are shown in Figure \ref{fig:08}b, and the peak value of $v_r^*$ decreases and then increases with the decrease in $R_t^*$. To investigate the influence of $R_t^*$ on the ring jet velocity $v_r^*$, we extract the gas-liquid interface at the moment of the counter jets meeting under different $R_t^*$, as shown in Figure \ref{fig:08}c. For large $R_t^*$, the bubble only occupies a part of the tube cross-section in the bubble contraction process, and the liquid on the two sides of the bubble can exchange momentum through the gap between the wall and the bubble. The viscous force near the wall produces an additional deceleration on the counter jets, making the fluid velocity near the wall remarkably lower than that near the tube axis. Hence, the gas-liquid interface has greater curvature at the axis. With the decrease in $R_t^*$, the bubble contraction time increases, and the deceleration effect is enhanced, resulting in a more significant increase in the interface curvature at the axis. In this case, greater curvature at the meeting moment of the counter jets means that less fluid is impinging and has a smaller total kinetic energy, and the resulting ring jet velocity decreases. However, when $R_t^*$ is reduced to a sufficiently small value ($R_t^* <  0.677$, as in Figure \ref{fig:08}c), the bubble can completely occupy the whole cross-section of the tube. The fluid inside the tube is sufficiently expelled so that the deceleration of the counter jets by the liquid close to the wall is small in the contraction stage of the bubble. The resulting counter jets have a larger diameter (as illustrated in Figure \ref{fig:08}c) and, therefore, a large amount of kinetic energy. In this case, the ring jet produced by the meeting of the counter jets has a high velocity.

\subsection{Effect of tube length}\label{sec:3.3}
\begin{figure*}
  \centering
  \includegraphics[scale=0.75]{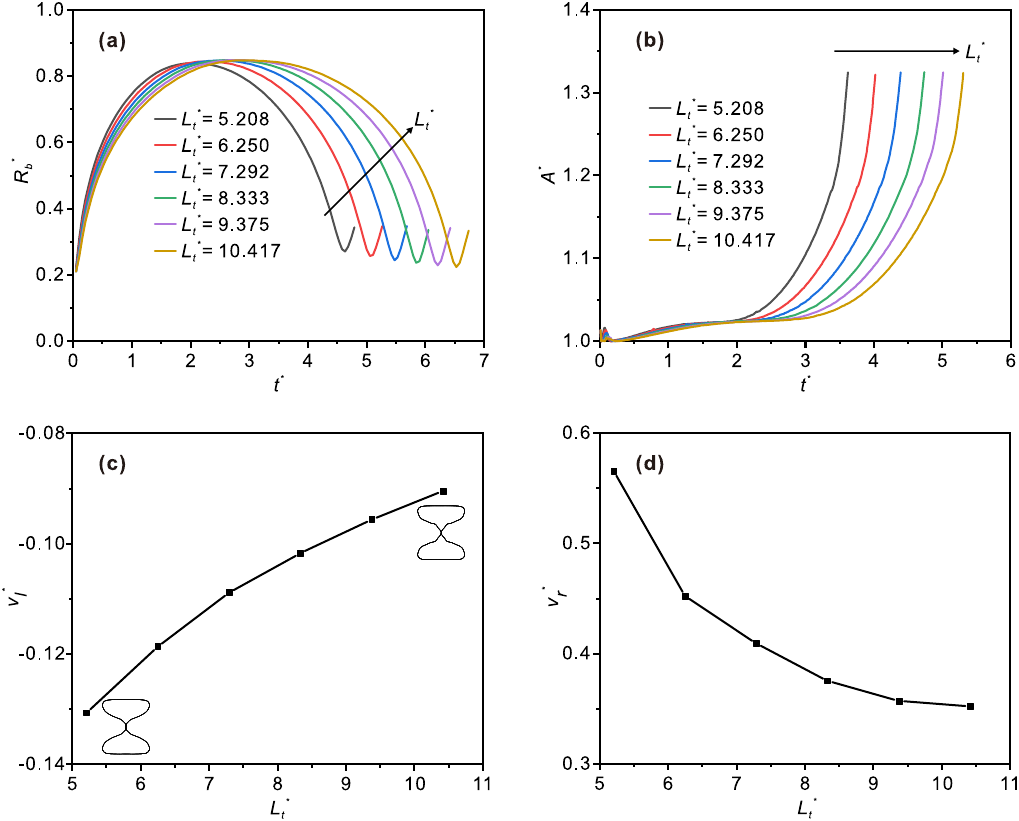}
  \caption{Influence of the tube length parameter $L_t^*$ on the bubble evolution with $R_t^*$ = 0.781, $\gamma = 0$: (a) variation of the bubble size $R_b^*$; (b) variation of the deformation parameter $A^*$; (c) jet velocity $v_l^*$ at the meeting moment of the counter jets; (d) peak velocity $v_r^*$ of the ring jet.}
  \label{fig:09}
\end{figure*}

The influence of the tube length on the bubble evolution is shown in Figure \ref{fig:09}. Figures \ref{fig:09}a-b present the size and the deformation of the bubble for different tube lengths. With the increase in  $L_t^*$, the maximum bubble size is almost unchanged and the bubble lifetime increases significantly. The increase in $L_t^*$ enhances the confinement effect of the tube wall on the bubble. This effect slows down the expansion and contraction of the bubble, which increases the bubble lifetime. We found that this conclusion still works for long tubes by simulating a tube with a large length parameter($L_t^*$ = 43.75). Even if the tube is very long, the change in tube length still has a significant effect on bubble lifetime. The deformation parameter $A^*$ at the moment of penetration of the ring jet does not differ much for different $L_t^*$, indicating that $L_t^*$ has little influence on the final degree of bubble deformation.

The variations of the counter jet velocity $v_l^*$ and the ring jet velocity $v_r^*$ with tube length are shown in Figures \ref{fig:09}c-d. The magnitude of $v_l^*$ decreases with the increase in the tube length during contraction, as shown in Figure \ref{fig:09}c (note that the velocity value is negative in Figure \ref{fig:09}c). A larger tube length parameter $L_t^*$ increases the wall confinement effect on the flow of fluid inside the tube. At the same time, the volume of fluid inside the tube increases, and more fluid volume has a greater inertia. The increase in the liquid inertia restricts the formation and development of the counter jets, and decreases the velocity of the counter jets. A larger $L_t^*$ prolongs the bubble contraction time, resulting in a greater deceleration effect of the lower velocity liquid near the tube wall on the counter jets. This increases the interface curvature at the axis before the counter jets meet, which finally decreases the peak velocity of the ring jet generated by the meeting of the counter jets, as shown in Figure \ref{fig:09}d.

\subsection{Effect of liquid viscosity}\label{sec:3.4}
\begin{figure*}
  \centering
  \includegraphics[scale=0.75]{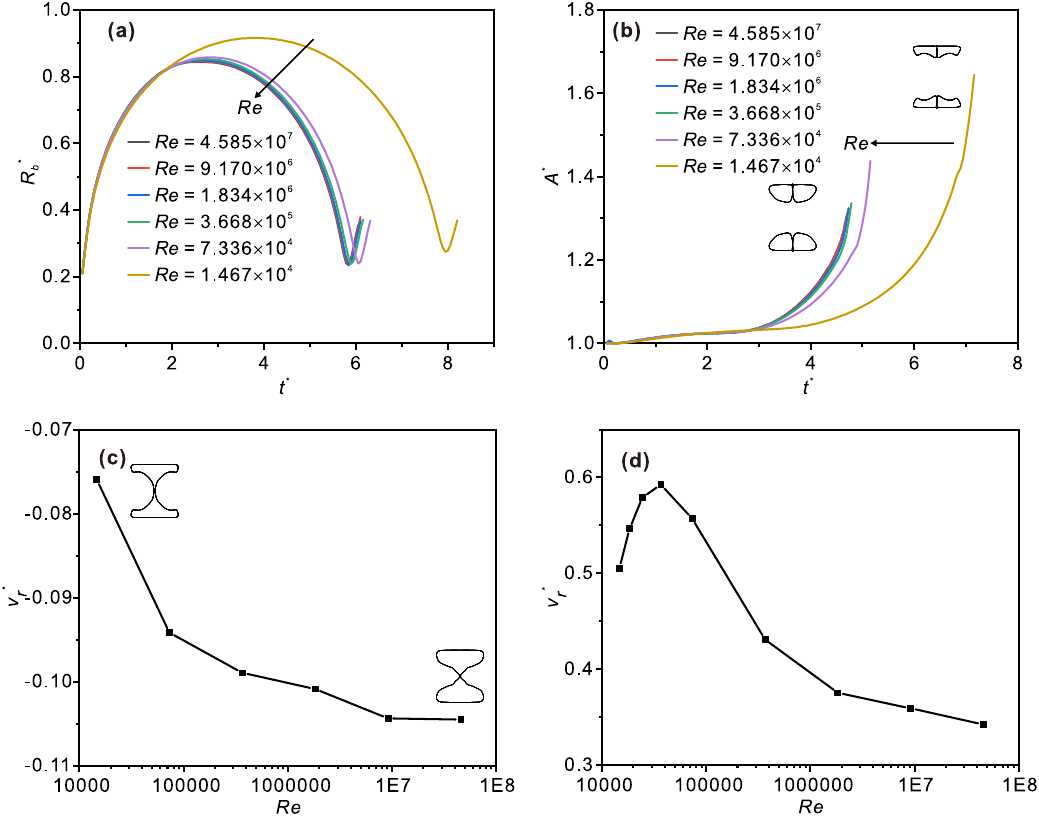}
  \caption{Influence of the Reynolds number $Re$ on the bubble evolution with $R_t^*$ = 0.781, $L_t^*$  = 8.333, $\gamma = 0$: (a) variation of the bubble size $R_b^*$; (b) variation of the deformation parameter $A^*$; (c) jet velocity $v_l^*$ at the meeting moment of the counter jets; (d) peak velocity $v_r^*$ of the ring jet.}
  \label{fig:10}
\end{figure*}

To study the influence of liquid viscosity, we use the Reynolds number $Re$, which can be defined by using the tube diameter as the characteristic length.
\begin{equation}\label{eq:19}
Re=\frac{2{{\rho }_{l}}{{v}_{m}}{{R}_{t}}}{{{\mu }_{l}}}.
\end{equation}
The variations of bubble size $R_b^*$ and deformation parameter $A^*$ with time $t^*$ for different Reynolds numbers $Re$ are shown in Figure \ref{fig:10}a-b. With the decrease of $Re$, the maximum size and lifetime of the bubble remain almost constant and then increase. This is because when $Re$ is large, the influence of the viscous force on the bubble evolution is negligible, the decrease of $Re$ has little effect on the bubble. When $Re$ is small enough, the effect of viscous force on bubble evolution is not negligible. With the decrease of $Re$, both the bubble expansion process and the contraction process are restricted by the viscous force. Since the pressure difference driving the bubble contraction is significantly smaller than that driving the bubble expansion, the bubble contraction process is more restricted by the increase in the viscous force. As a result, the maximum size of the bubble increases. In addition, the increase in viscous forces inhibits the development of the counter jets, which leads to an increase in the diameter of the counter jets. This increases the degree of deformation of the ring bubble produced by the meeting of the counter jets.

The variations of the counter jet velocity $v_l^*$ and the ring jet velocity $v_r^*$ with $Re$ are shown in Figures \ref{fig:10}c-d, respectively. The magnitude of $v_l^*$ decreases with the decrease in $Re$ during contraction, as shown in Figure \ref{fig:10}c (note that the velocity value is negative in Figure \ref{fig:10}c). This is because as $Re$ decreases, the confinement effect on the counter jets by the viscous forces is enhanced. With the decrease of $Re$, the peak velocity of the ring jet $v_r^*$ increases and then decreases. This is because when $Re$ is large ($Re > 36680$), with the decrease of $Re$, the viscous effect on the counter jets is enhanced, and the curvature of the gas-liquid interface decreases at $r$ = 0 before the counter jets meet, which finally increases the peak velocity of the ring jet generated by the meeting of the counter jets. When $Re$ is small ($Re < 36680$), with the decrease of $Re$, the enhancement of the viscous force inhibits the development of the ring jet, resulting in the decrease in the peak velocity of the ring jet.

\subsection{Effect of initial bubble position}\label{sec:3.5}
\begin{figure}
  \centering
  \includegraphics[scale=0.8]{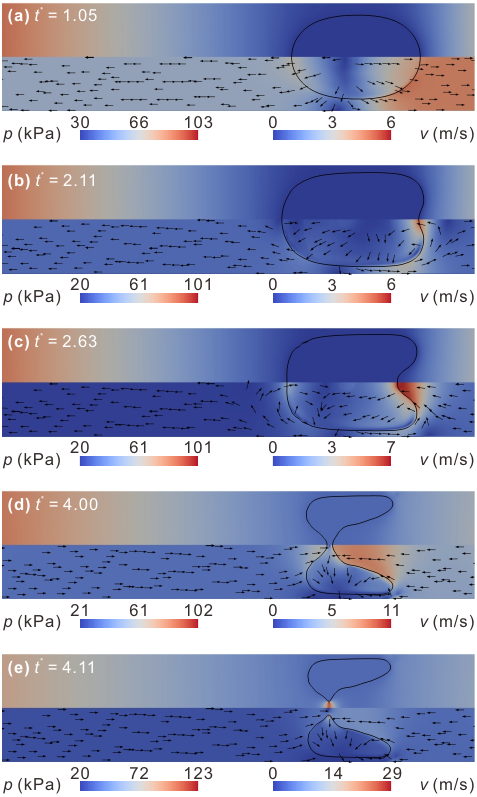}
  \caption{Bubble shape as well as pressure and velocity distributions during bubble evolution for the case of the bubble deviating from the midpoint of the tube. Here, $R_t^* = 0.781$, $L_t^* = 8.333$, and $\gamma = 1.563 $ (Multimedia available online).}
  \label{fig:11}
\end{figure}

As suggested by the results in Section \ref{sec:3.3}, the distance between the bubble and the tube end affects the bubble evolution process, so we analyze the bubble dynamics when the bubble deviates from the midpoint of the tube in the axial direction. A typical case (with dimensionless offset parameter $\gamma = 1.563 $) is presented in Figure \ref{fig:11} (Multimedia available online). In the early expansion stage, the right gas-liquid interface grows significantly faster than the left because the amount of liquid on the right is less than that on the left, as shown in Figure \ref{fig:11}a. As the bubble expands, the bubble pressure becomes lower than the ambient pressure, which leads to the bubble expansion rate decreasing continuously. Due to the less liquid on the right side, the gas-liquid interface velocity on the right side first decreases to zero and then begins to contract; at this time, the bubble's left interface is still expanding, as shown in Figure \ref{fig:11}b. At the time that the left interface starts to contract, the right interface already has a large contraction velocity, as shown in Figure \ref{fig:11}c. During the contraction process, the speed of the right interface is higher than that of the left until the two counter jets meet, as shown in Figure \ref{fig:11}d. The meeting of the two counter jets penetrates the bubble, and generating high-pressure at the point where the counter jets meet, forming a ring jet. The meeting location of the counter jets is closer to the midpoint of the tube than the initial bubble position, as shown in Figure \ref{fig:11}e.

\begin{figure*}
  \centering
  \includegraphics[scale=0.75]{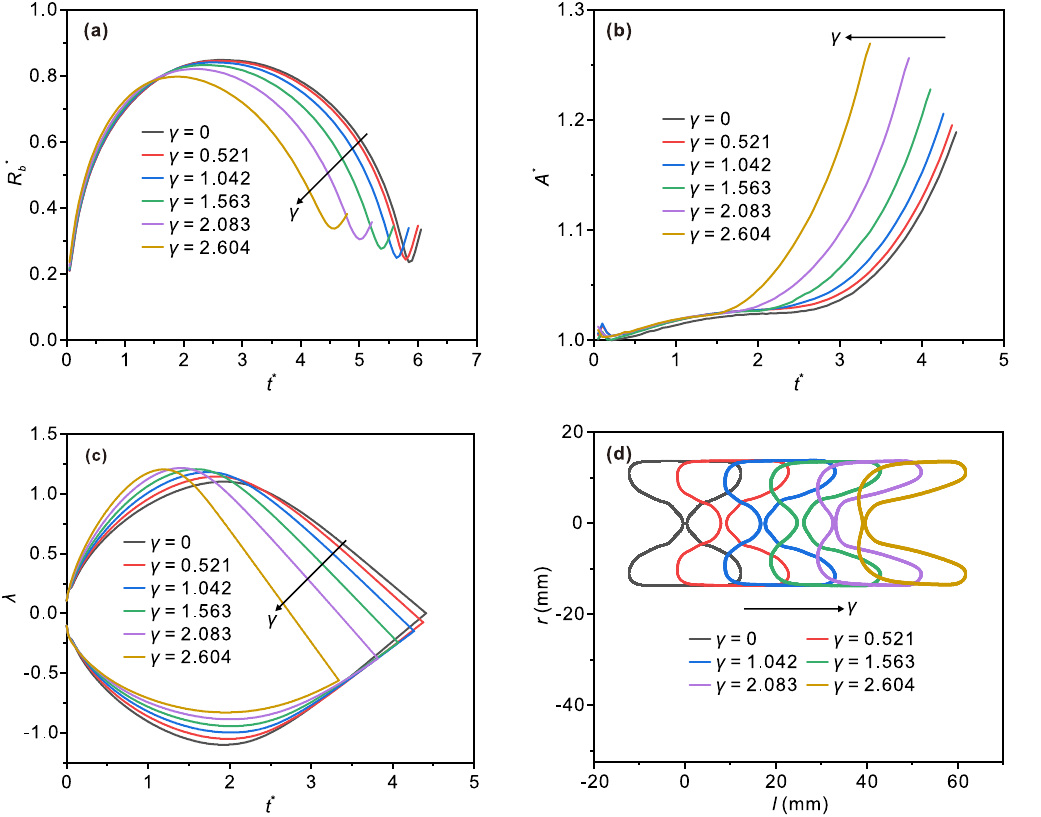}
  \caption{Influence of the dimensionless initial offset $\gamma $ on bubble evolution with $R_t^* = 0.781$, $L_t^* = 8.333$: (a) variation of the bubble size $R_b^*$; (b) variation of the deformation parameter $A^*$; (c) variation of gas-liquid interface positions on the two sides of the bubble; (d) bubble shape at the meeting moment of the counter jets.}
  \label{fig:12}
\end{figure*}

The impact of the dimensionless initial offset $\gamma $ on the bubble size and deformation are shown in Figures \ref{fig:12}a and \ref{fig:12}b, respectively. With the increase in the dimensionless initial offset $\gamma $, the bubble is closer to the right end of the tube, and the difference in the volumes of liquid between the two sides of the bubble increases. In this case, the bubble volume reaches its maximum value when the expansion of the bubble's left side and the contraction of the bubble's right side cancel each other's effect on the bubble volume. Compared with the expansion velocity of the bubble's left side, the contraction velocity of the bubble's right side is significantly larger, as shown in Figure \ref{fig:11}b. This leads to a decrease in the maximum bubble volume. With the increase of the offset parameter $\gamma $, the difference between the velocities on the bubble's two sides increases, which leads to a decrease in the bubble's maximum volume and lifetime, as shown in Figure \ref{fig:12}a. With increasing $\gamma $, the asymmetry in the axial direction increases, increasing the deformation parameter $A^*$, as shown in Figure \ref{fig:12}b.

To characterize the degree of bubble migration towards the tube midpoint during evolution, we define the dimensionless gas-liquid interface position
\begin{equation}\label{eq:20}
\lambda =\frac{{{x}_{\text{int}}}}{{{R}_{m}}},
\end{equation}
where $x_\text{int}$ is the instantaneous axial coordinate of the bubble's interfaces on the two sides relative to the initial bubble position.

To characterize the impact of the initial offset $\gamma $ on the bubble migration, we analyze the variation of the bubble interfacial positions with time for different dimensionless initial offsets $\gamma $, as shown in Figure \ref{fig:12}c. In the figure, the upper half of the curve is the interface position near the tube end (i.e., the right interface), the lower half is that near the midpoint (i.e., the left interface), and the intersection point of the two curves is the position where the counter jets meet. If the dimensionless interface positions on the left and right are equal in magnitude and opposite in sign (as shown by $\gamma = 0$ in Figure \ref{fig:12}c), it indicates that the bubble does not migrate during the evolution process. In contrast, if the magnitudes of the position parameters of the left and right interfaces are different, it indicates that the bubble migrates in the axial direction. As shown in Figure \ref{fig:12}c, the right interface velocity is significantly higher than the left, which is because of the difference in the liquid amount on the two sides of the bubble, i.e., decreasing the liquid amount can accelerate the bubble interface movement. Therefore, with the increase in the dimensionless initial offset $\gamma$, the degree of axial migration of the bubble increases.

To consider the impact of the dimensionless initial offset on the variation of the gas-liquid interface, we extract the interface of the bubble at the meeting time of the counter jets, as shown in Figure \ref{fig:12}d. The interfaces on the two sides of the bubble are symmetric for $\gamma = 0$. With increasing $\gamma $, the difference in the interface shape between the two sides becomes larger: the left interface (e.g., close to the midpoint) is smoother, and the right interface (e.g., close to the tube end) is more concave. This further verifies that the initial offset leads to differences in the contraction of the left and right interfaces and differences in the two counter jets.

From Sections \ref{sec:3.2} and \ref{sec:3.3}, we can see that the tube diameter and length affect the expansion/contraction of the bubble, so we analyze the influence of the tube diameter and length on the bubble migration in the case where the bubble deviates along the axial direction.

\begin{figure*}
  \centering
  \includegraphics[scale=0.75]{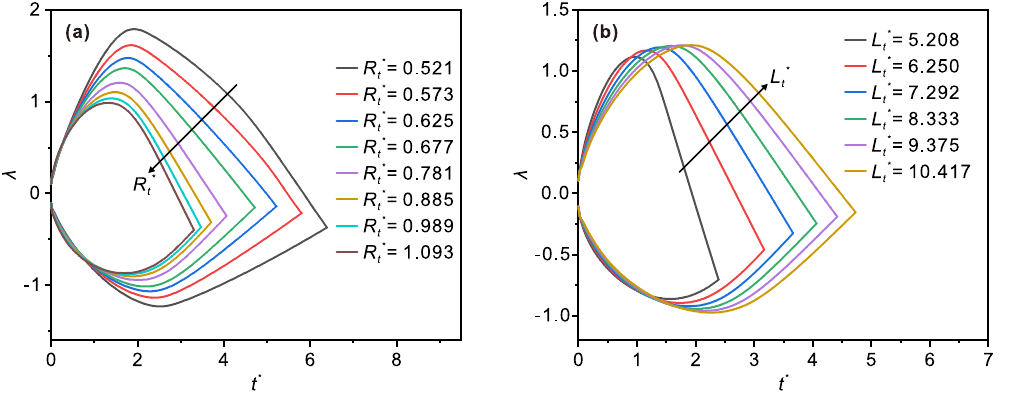}
  \caption{(a) Influence of the tube diameter $R_t^*$ on the position of the gas-liquid interface with $L_t^* = 8.333$, $\gamma = 1.563 $. (b) Influence of the tube length $L_t^*$ on the position of the gas-liquid interface with $R_t^* = 0.781$, $\gamma = 1.563 $. }
  \label{fig:13}
\end{figure*}

The variations of the interface positions for different $R_t^*$ are shown in Figure \ref{fig:13}a. With the decrease in the tube diameter R, the degree of bubble migration first decreases and then increases. At large $R_t^*$, the bubble fails to fully occupy the tube cross-section, and the decrease in $R_t^*$ can strengthen the confinement of the wall, inhibiting the development of the interface velocities. Because of less liquid on the right, the inertia of the liquid is smaller, and the degree of decreasing interface velocity of the bubble on this side is greater, reducing the difference between the left and right interface velocities, and thus the degree of migration is decreased. If $R_t^*$ decreases so that the bubble completely occupies the tube cross-section, it is difficult for the liquid on the two sides of the bubble to pass through the gap between the bubble and the tube wall in the contraction process. This leads to the right interface maintaining a higher velocity, while the left interface velocity continues to decrease due to the increased wall restriction. This causes an increase in the difference between the left and right interfacial velocities and also an increase in the degree of bubble migration.
The variations in the interface positions on the two sides of the bubble for different $L_t^*$ are shown in Figure \ref{fig:13}b. The degree of bubble migration decreases with the increase in $L_t^*$. The increase in $L_t^*$ causes the liquid volume on the two sides of the bubble to increase, and the increase in liquid inertia reduces in the interface velocities on the two sides. However, because the relative difference between the liquid volume on the two sides of the bubble decreases, the difference between the interface velocities on the two sides of the bubble decreases, and the degree of migration decreases.

\section{Conclusions}
In summary, we numerically investigate a cavitation bubble in a thin tube, focusing on the confinement effects of the tube on the counter jets, ring jets, and axial migration of the bubble. For bubbles initially at the midpoint of the tube, the bubble evolution is symmetric, in which the tube diameter parameter $R_t^*$, the tube length parameter $L_t^*$, and the Reynolds number $Re$ have a strong influence on the degree of bubble confinement effects of the tube. The restriction effect by the tube increases with the decrease in the tube diameter parameter $R_t^*$, the bubble life is prolonged, the deformation degree increases, and the counter jet velocity decreases. In particular, when $R_t^*$ is small enough, the tube wall restricts the bubble so strongly that the gap between the wall and the bubble is too narrow, resulting in a high ring jet velocity. With increasing the tube length parameter $L_t^*$, the restriction effect by the wall increases, the lifetime of the bubble is prolonged, and the velocities of the counter jets and the ring jet both decrease. With decreasing of the Reynolds number $Re$, the resistance to flow in the tube increases, the bubble lifetime prolongs, the maximum size increases, and the counter jet velocity decreases. If the bubble deviates from the midpoint of the tube initially, the two sides of the bubble show different behaviors. The gas-liquid interface on the side near the tube end expands faster and starts to contract earlier than the side near the midpoint of the tube, and contracts faster until the counter jets meet. The difference between the left and right interfaces leads to the migration of the bubble toward the midpoint of the tube. With the increase in $\gamma $, the degree of the asymmetric confinement effects of the tube increases, and the degree of bubble migration increases. The tube diameter parameter $R_t^*$ and the tube length parameter $L_t^*$ also have large influences on the bubble migration. The findings about the deformation/migration of the bubble and formation of jets in this study not only unveil the confinement effect of thin tubes on bubble dynamics, but also further demonstrate the complexity of the cavitation process in limited spaces. In addition, this study also suggests the possibility of controlling bubble dynamics and jet dynamics by configuring the geometry of confinement in real applications, such as fluid engineering, microfluidics, and clinical medicine. This study only considers the confinement effect of thin rigid tubes, and the effects of other confinement shapes and more details of the interactions between the bubble and the confinements are yet to be explored.

\section*{Acknowledgements}
This work is supported by the National Natural Science Foundation of China (Grant nos.\ 51920105010, and 51921004).

\section*{Data Availability Statement}
The data that support the findings of this study are available from the corresponding author upon reasonable request.

\section*{References}
\bibliography{cavitationBubble}
\end{document}